\documentstyle[epsfig, twocolumn]{venice97}

\begin{document}

\setlength{\parindent}{0pt}
\setlength{\parskip}{ 10pt plus 1pt minus 1pt}
\setlength{\hoffset}{-1.5truecm}
\setlength{\textwidth}{17.1truecm}
\setlength{\columnsep}{1truecm}
\setlength{\columnseprule}{0pt}
\setlength{\headheight}{12pt}
\setlength{\headsep}{20pt}
\pagestyle{veniceheadings}

\title{\bf HIPPARCOS PARALLAXES AND PROPER MOTIONS OF CEPHEIDS AND THEIR
IMPLICATIONS\thanks{Based on data from the Hipparcos astrometry satellite}}

\author{{\bf M.W.~Feast$^1$, P.A.~Whitelock$^2$} \vspace{2mm}\\
$^1$Astronomy Department, University of Cape Town, Rondebosch, 7700, South
Africa\\
$^2$South African Astronomical Observatory, PO Box 9, Observatory, 7935,
South Africa}

\maketitle

\begin{abstract}
 A summary is given of an analysis of the Hipparcos trigonometrical
parallaxes and proper motions of classical Cepheids. It is possible for the
first time to derive zero-points for the period-luminosity and
period-luminosity-colour relations from parallaxes alone, avoiding the
problems of less direct methods. The results imply an increase of 8 to 10
percent in the extragalactic distance scale based on Cepheids. The proper
motions are used to derive the constants of galactic rotation. Comparison
with radial velocity data leads to a confirmation of the Cepheid distance
scale derived from the parallaxes and indicates a kinematic distance to the
galactic centre of $\rm 8.5 \pm 0.5\: kpc$. From the new Cepheid distances
to the LMC and M31, the absolute magnitude of RR Lyrae variables in
metal-poor globular clusters is derived. Applying this to data on metal-poor
clusters in our own Galaxy leads to an age of about 11 Gyr for these
clusters, considerably less than previously thought.
Other evidence from Hipparcos on these matters is briefly reviewed and it is
suggested that the Cepheid results currently provide the most reliable scale
on which to base distances and ages.\vspace {5pt}\\

Key~words: space astrometry; Cepheid variables; distance scale; galactic 
structure; globular clusters.\\

\end{abstract}

\section{INTRODUCTION}
The measurement of the trigonometrical parallaxes and proper motions of
Cepheid variables by the Hipparcos satellite (ESA 1997) represents a major
advance over earlier work because of the scope and accuracy of the new results
and the fact that they are referred to a co-ordinate system which is
uniform over the whole sky and based on the positions of distant extragalactic
objects (see ESA 1997 and other papers in this present volume). In the present
paper we summarise results obtained by Feast \& Catchpole (1997) (=FC) and
by Feast \& Whitelock (1997) (=FW) from analyses of the Hipparcos Cepheid 
data-base. We discuss the implications of these results for galactic
structure, for the extragalactic distance scale and for the ages of the
oldest stars. Finally we make a brief comparison with other Hipparcos results
which have implications for distance scales and stellar ages.

\section{THE ZERO-POINT OF THE CEPHEID PL AND PLC RELATIONS}
The determinations of the zero-points of the Cepheid period-luminosity (PL)
and period-luminosity-colour (PLC) relations are essential keys to
galactic and extragalactic problems.\\

Until the Hipparcos data became available the most satisfactory way of
obtaining these zero-points was from Cepheids in open clusters in our
Galaxy whose distances could be estimated from main-sequence fitting
procedures. A  problem with this method is that, because of the steepness
of the main-sequence, it is quite sensitive to the adopted reddening and to
the accuracy of the photometry. In addition the Hipparcos results themselves
(van Leeuwen \& Hansen Ruiz 1997, Mermilliod et al.\ 1997) suggest that there
may be previously unsuspected complications with the main-sequence fitting
method itself.\\

It is now possible to estimate the PL and PLC zero-points with good
accuracy, directly from the Hipparcos trigonometrical parallaxes. In the
first distribution of Hipparcos data there are trigonometrical parallaxes
for just over 200 classical Cepheids. Since Cepheids are mostly distant,
even for Hipparcos, most of the parallaxes are small and some, of course,
are negative. It is thus exceptionally important that the data are combined
in a way that avoids statistical bias effects. Most usual ways of combining
the results will in fact be subject to large bias effects of the general
Lutz-Kelker type and the results obtained in these ways will lead to an
underestimate of the luminosity of the Cepheids. However, unbiased results
can be obtained once we accept that we are not attempting to use the
parallaxes to determine the distances or absolute magnitudes of individual
Cepheids but simply to determine the zero-points of the PL and PLC
relations. These relations (without their zero-points) allow accurate
relative distances of Cepheids to be established and a correct analysis of
the data then enables one to reduce any bias to negligible proportions.\\

The PLC relation,
\begin{equation}
\langle M_{V}\rangle  = \alpha \log P + \beta (\langle B\rangle -\langle V\rangle
)_{0} + \gamma,
\end{equation}
 has negligible intrinsic scatter. One needs however, to have estimates of
individual reddenings in order to use it. These can be obtained from
multicolour photometry and in the present work we have used estimates from
$BVI$ photometry.

The PL and PC relations,
\begin{equation}
\langle M_{V}\rangle  = \delta \log P + \rho
\end{equation}
\begin{equation}
(\langle B\rangle -\langle V\rangle )_{0} =\tau \log P + \theta, 
\end{equation}
 both have significant intrinsic scatter since they are both approximations
to the PLC relation. However, the scatter in the two equations is correlated
and, since $ A_{V}/E_{B-V} \sim \beta$, an error in deriving $ A_{V}$ from
Equation 3 is largely compensated for when applied in the use of Equation 2.
In this way the effective scatter in the PL relation is reduced by a factor
of about three.\\

In our work we adopt a galactic PC relation from Laney \& Stobie (1994), viz.;
\begin{equation}
(\langle B\rangle -\langle V\rangle )_{0} = 0.416 \log P + 0.314,
\end{equation}
and the slope of the PL relation as determined by Caldwell \& Laney (1991) in
the LMC,
\begin{equation}
\langle M_{V}\rangle  = -2.81 \log P + \rho .
\end{equation}
Then an essentially bias free estimate of $\rho$ can be obtained from
a properly weighted solution to the following equation.
\begin{equation}
10^{0.2\rho} = 0.01\, \pi\, 10^{0.2(\langle V_{0}\rangle  + 2.81 \log P)}
\end{equation}
where $\pi$ is the Hipparcos parallax in mas. The weighting depends
essentially on the reciprocal  of the square of the standard error
of the parallax ($\sigma_{\pi}$, see FC for details).
It should be particularly noticed that there is no selection or weighting
by $\pi$ or $\sigma_{\pi}/\pi$ and negative parallaxes can be included.
Any subdivision of the data by $\pi$ or $\sigma_{\pi}/\pi$ inevitably
leads to biased results.\\

There is a further slight complication in that some of the Hipparcos
Cepheids are known to be overtone pulsators and in those cases the fundamental
period has been used (see FC).\\

In the full sample there are 220 Cepheids. But if we omit Polaris, 75
percent of the weight is in 25 stars. The table shows some typical solutions
for $\rho$ (further solutions are in FC). N is the number of Cepheids in
each solution Solutions have been done with and without Polaris. In fact the
parallax of Polaris has sufficient weight that it can be used alone as in
the table. A comparison with the general solutions shows rather clearly that
Polaris is an overtone pulsator. This is not really surprising in view of
its low amplitude. However, it has, in fact, been used in the past with its
companion as a calibrating Cepheid assuming it to be a fundamental
pulsator.\\

\begin{table}[htb]
\caption{\em Some results for the zero-point, $\rho$, of the Cepheid PL relation
from Hipparcos parallaxes}
\begin{center}
\leavevmode
\footnotesize
\begin{tabular}[h]{lrc}
    \hline \\ [-5pt]
Solution                    &   N   &       $\rho$       \\[+5pt]
\hline \\[-5pt]
Whole sample       & 220 & $-1.40 \pm 0.12$ \\
$\alpha$ UMi (fundamental) & 1 & $-2.04 \pm 0.14$ \\
$\alpha$ UMi (first overtone) & 1 & $-1.41 \pm 0.14$ \\
$\alpha$ UMi (second overtone) & 1 & $-0.97 \pm 0.14$ \\
Whole sample less $\alpha$ UMi & 219 & $-1.40 \pm 0.16$ \\
High weight less $\alpha$ UMi &  25 &  $-1.44 \pm 0.13$ \\
High weight plus $\alpha$ UMi &  26 &  $-1.43 \pm 0.10$ \\
\hline \\
\end{tabular}
\end{center}
\end{table}

From these results we adopt for the PL zero-point,\\
$\rho = -1.43 \pm 0.10$.\\
A similar procedure using $BVI$ reddenings gives a zero-point for the
PLC relation (Equation 1) of,\\
$\gamma = -2.38 \pm 0.10$,\\
where we have adopted $\alpha = -3.80$ and $\beta = +2.70$ (see FW).

\begin{figure*}
\centering
\epsffile[-7 85 425 340]{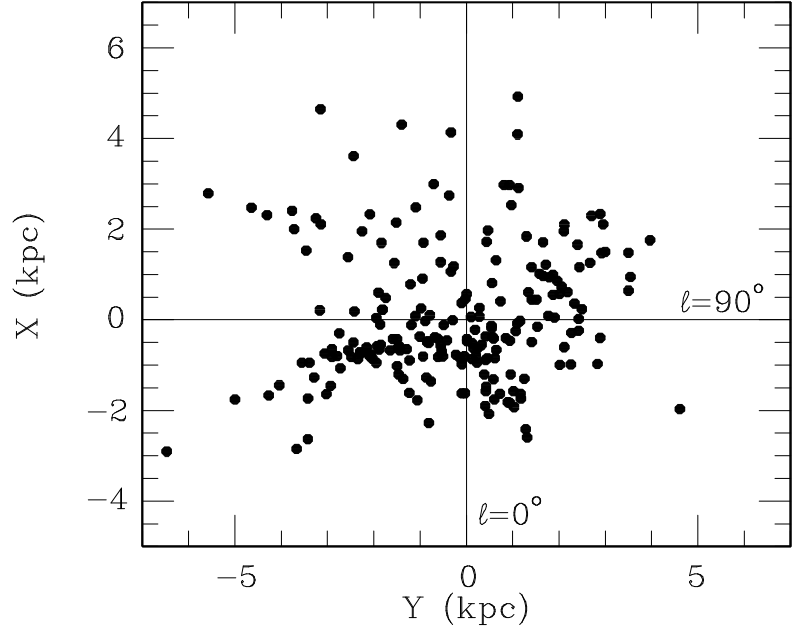}
 \caption{\em The distribution of the Cepheids used in the proper motion
solutions seen projected onto the galactic plane.  The Sun is at the origin
of the co-ordinate system. The distances are from the PL relation derived in
FC}
 \end{figure*}

\section{HIPPARCOS PROPER MOTIONS OF CEPHEIDS} The Cepheids with Hipparcos
proper motions extend over a region of several kiloparsecs in the galactic
plane around the Sun (Figure 1). They are therefore important for the study
of galactic kinematics and for a comparison of the Oort constant, $A$, with
that obtained from radial velocities. Since the value of $A$ derived from
radial velocities depends on the distance scale adopted whilst that from
proper motions is (nearly) independent of the distance scale, a comparison
of the two allows an estimate of this scale to be made. This may then be
compared with that obtained from the trigonometrical parallaxes.\\

Figure 2 shows the proper motion in galactic longitude, multiplied by
$\kappa = 4.74047$, and with the local solar motion subtracted, plotted
against galactic longitude. The effects of galactic rotation are very
clearly seen. The curve is an Oort-type first order solution. The three
stars in the plot that lie well away from the mean curve are nearby and their
proper motions are dominated by their peculiar velocities. They have little
weight in the solutions. Solutions for galactic rotation have been made
introducing higher order terms in the equations as in the radial velocity
study of Cepheids by Pont et al.\ (1994).  Full details are given by FW.

\begin{figure*}
\centering
\epsffile[7 85 425 340]{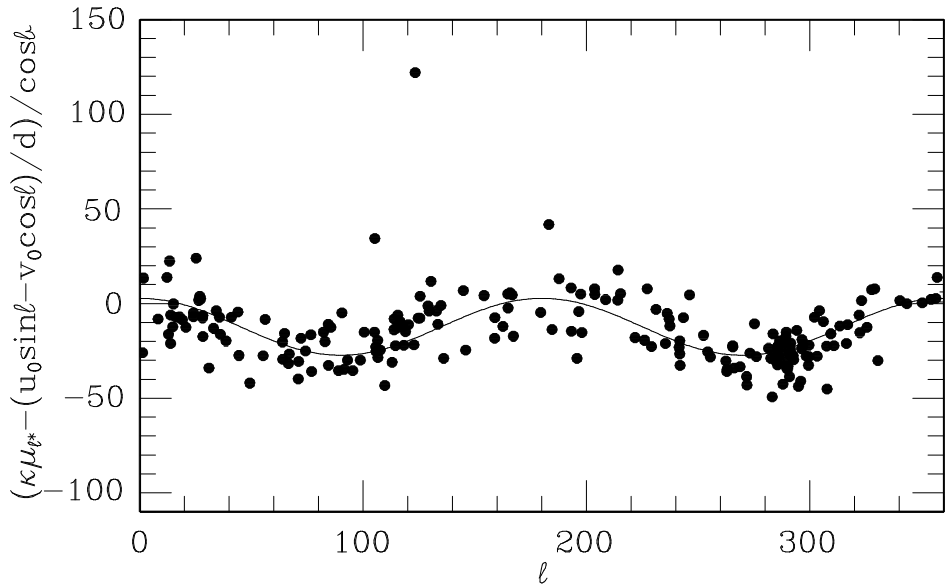}
 \caption{\em The proper motion in galactic longitude multiplied by $\kappa$,
and corrected for local solar motion, plotted against galactic longitude.
The curve corresponds to solutions discussed in the text. The three
outstanding stars are nearby and have low weight in the solution.}
 \end{figure*}

The principal results are;\\
$ A = 14.82 \pm 0.84\rm \: km\, s^{-1} kpc^{-1}$\\
$ B = -12.37 \pm 0.64\rm \: km\, s^{-1} kpc^{-1}$\\
$ \Omega_{\rm o} = 27.19 \pm 0.87 \rm \: km\, s^{-1} kpc^{-1}$\\
$(d\Theta/dR)_{\rm o} = -2.4 \pm 1.2\rm \: km\, s^{-1} kpc^{-1}$\\
where $A$ and $B$ are the usual Oort constants,
$\rm \Omega_{\rm o}$ is the angular velocity of circular rotation
at the Sun, $\rm \Theta$ is the circular velocity and $R$ the distance
from the galactic centre. Thus the negative $\rm (d\Theta/dR)_{\rm o}$
indicates that the velocity of galactic rotation is declining (slowly)
at the solar distance from the galactic centre. A comparison with the
value of $A$ derived by Pont et al.\ (1994), who used a PLC relation
to derive Cepheid distances, leads to a PLC zero-point,\\
$\gamma = -2.42 \pm 0.13.$\\
This is close to that derived above from the trigonometrical parallaxes, \\
$\gamma = -2.38 \pm 0.10.$\\
Adopting the parallax zero-point the distance to the galactic centre
derived kinematically from Cepheid radial velocities by Pont et al.\
changes from the value they give to:\\
$ R_{\rm o} = 8.5 \pm 0.5 \rm \: kpc$\\

\section{CONSEQUENCES OF THE HIPPARCOS CEPHEID DISTANCE SCALE}
  If we  use the PL relation for the LMC (Caldwell \& Laney 1991) with the
Hipparcos zero-point and a small metallicity correction (Laney \& Stobie
1994) we obtain an true LMC distance modulus of 18.70 mag. This is a 10
percent increase in distance over that used for the basic distance scale in
the HST work on the extragalactic distance scale based on Cepheids (e.g.\
Freedman et al.\ 1994). If we adopt Gould's (1994) relative distance of M31
and the LMC then we get a 17 percent increase in distance for M31 compared
with the value derived by Freedman \& Madore (1990).\\

There is a point here that seems to have caused some confusion.  If one
compares the Hipparcos PL relation,\\
\begin{equation}
\langle M_{V}\rangle  = -2.81 \log P -1.43
\end{equation}
with that used in the HST distance-scale work,\\
\begin{equation}
\langle M_{V}\rangle  = -2.76 \log P -1.40, \end{equation}
 one sees that there is only about 4 percent difference in the two distance
scales in the range of periods of interest. However, this leaves out the
fact that it is essential in using the Hipparcos scale to also use a
reddening scale with the same zero-point as that use with the Hipparcos
calibrating Cepheids. The HST workers adopt, rather arbitrarily, a mean LMC
reddening of $E_{B-V} =0.1$. The scale used in FC with the Hipparcos
calibrating Cepheids leads to a mean reddening for the relevant LMC Cepheids
of $E_{B-V} = 0.076$. Taking this difference into account leads finally to a
difference between the Hipparcos and HST scales of 8 percent. This is not
significantly different from the 10 percent estimated above.\\

The Hipparcos Cepheid scale allows us to estimate the absolute magnitudes of
RR Lyrae stars in metal-poor globular clusters in the LMC and in M31. We
find that $M_{V}(\rm RR) = 0.3$ mag at $\rm [Fe/H] =-1.9$. This is 0.3 mag
brighter than has recently been assumed in estimates of the ages of galactic
metal-poor globular clusters and leads to a significant reduction in the
ages of these clusters (to about 11 Gyr). It should be noticed that the
strength of this method is that it involves a direct comparison of globular
clusters of the same metallicity in the various galaxies.\\

This result was perhaps surprising in view of the fact that pre-Hipparcos
statistical parallaxes of RR Lyrae variables suggested fainter absolute
magnitudes ($M_{V}({\rm RR}) = 0.71 \pm 0.12$ at $[Fe/H] = -1.6$) and the
Hipparcos proper motions lead to rather similar results (Fernley et al.\
1997). However, the statistical parallax solutions are weighted to higher
metallicities than the clusters and they refer to field RR Lyraes rather
than directly to cluster stars. In addition it is important to remember that
in the statistical parallax analysis a model is required for the galactic
halo. Generally a rather simple model is adopted. However, we are beginning
to realize that the galactic halo is in fact quite complex and may be
dominated by the remnants of a few infalling satellite galaxies.\\

Hipparcos potentially offers another line of enquiry regarding these problems.
This is to use the
parallaxes of metal-poor subdwarfs to derive distances 
(and hence ages) of globular
clusters from main-sequence fitting techniques. Reid (1997) has carried
out such an analysis and his results for the distance of the LMC and the
age of the oldest globular clusters are in remarkably close agreement
with those derived above from the Hipparcos Cepheid scale. Rather similar
results have been obtained from Hipparcos subdwarfs by Gratton et al.\ (1997) 
(see also
Fusi Pecci et al.\ 1997). On the other hand Pont et al.\ (1997) report 
that they obtain a much shorter distance scale from the Hipparcos subdwarf
data. It is important that this conflict between the different groups
analysing the Hipparcos subdwarf data be clarified. The difference 
between the groups seems
to be in the manner in which statistical bias is treated. It should
therefore
be borne in mind that corrections for statistical bias are (of their
nature) often quite uncertain (see, e.g.\ Koen  1992) and 
that the Hipparcos Cepheid scale discussed above is free of these
problems.\\

\section{CONCLUSIONS}
The Hipparcos trigonometrical parallaxes of Cepheids lead to:\\
1. An increase in the basic ``HST'' extragalactic distance scale of
8 to 10 percent.\\
2. A substantial decrease in the previously adopted ages of metal-poor
globular clusters (to about 11 Gyr).\\
3. An increase in the kinematic distance to the Galactic Centre 
(to $ R_{\rm o} = 8.5 \pm 0.5 \: {\rm kpc}$).\\

The Hipparcos proper motions of Cepheids;\\
1. Confirm the parallax scale.\\
2. Yield the following values:\\
  $A = 14.82 \pm 0.84\rm \: km\, s^{-1} kpc^{-1}$;\\
  $B = -12.37 \pm 0.64\rm \: km\, s^{-1} kpc^{-1}$;\\
  $\rm \Omega_{\rm o} = 27.19 \pm 0.87\: km\, s^{-1} kpc^{-1}$;\\
  $\rm (d\Theta/dR)_{\rm o} = -2.4 \pm 1.2\: km\, s^{-1} kpc^{-1}$.\\

A number of factors suggest that the new Cepheid distance scale discussed in
this paper is more robust than others that have been discussed recently.

\end{document}